\documentclass[aps,nofootinbib,preprint]{revtex4}
\usepackage{amsfonts,amssymb,amsmath}
\usepackage{graphicx}
\usepackage{booktabs}
\def\d{\delta}
\def\g{\sqrt{-g}}

\newcommand{\f}[2]{\frac{#1}{#2}}
\newcommand{\mk}[1]{\left( #1 \right)}
\newcommand{\kk}[1]{\left[ #1 \right]}

\newcommand{\be}{\begin{equation}}
\newcommand{\ee}{\end{equation}}
\newcommand{\DE}{{\rm DE}}
\newcommand{\vecs}[1]{\mbox{\boldmath\tiny ${#1}$}}
\newcommand{\veck}{\vecs k}
\def\eff{{\rm eff}}
\newcommand{\lcdm}{{\rm \Lambda CDM}}

\begin{document}%%%%%%%%%%%%%%%%%%%%%%%%%%%%%%%%%%%%%%%%%

\title{
$f(R)$ Gravity and its Cosmological Implications
}

\author{
Hayato Motohashi$^{~1,2}$, Alexei A. Starobinsky$^{~2,3}$,
and Jun'ichi Yokoyama$^{~2,4}$
}

\address{
$^{1}$ Department of Physics, Graduate School of Science,
The University of Tokyo, Tokyo 113-0033, Japan \\
$^{2}$ Research Center for the Early Universe (RESCEU),
Graduate School of Science, The University of Tokyo, Tokyo 113-0033, Japan \\
$^{3}$ L. D. Landau Institute for Theoretical Physics,
Moscow 119334, Russia \\
$^{4}$ Institute for the Physics and Mathematics of the Universe(IPMU),
The University of Tokyo, Kashiwa, Chiba, 277-8568, Japan
}

\begin{abstract}%%%%%%%%%%%%%%%%%%%%%%%%%%%%%%%%%%%%%%%%%
We have investigated the evolution of a homogeneous isotropic
background of the Universe and inhomogeneous subhorizon matter
density perturbations in viable $f(R)$ models of present dark
energy and cosmic acceleration analytically and numerically. It is
found that viable $f(R)$ models generically exhibit recent
crossing of the phantom boundary $w_{\rm DE}=-1$. Furthermore, it
is shown that the growth index of perturbations depends both on
time and wavenumber. This anomalous growth may explain properties
of the observational matter power spectrum from the SDSS data and
can also partially counteract the spectrum suppression by massive
neutrinos making larger values of the total sum of neutrino
rest masses possible.
\end{abstract}

\begin{flushright}
RESCEU-02/11
\end{flushright}

\maketitle

%%% Body %%%%%%%%%%%%%%%%%%%%%%%%%%%%%%%%%%%%%%

\section{Introduction}%%%%%%%%%%%%%%%%%%%%%%%%%%%%%%%%%%%%%%%%%

Although the standard spatially flat ${\rm
\Lambda}$-Cold-Dark-Matter ($\lcdm$) model can explain cosmic
acceleration and is consistent with 
current observational data\cite{WMAP7}, the observed value of the
cosmological constant term
is much smaller than any other
energy scale known in physics.
On the other hand, we are sure that ``primordial dark energy
(DE)'', which is responsible for inflation in the early universe
\cite{S80,sato,guth}, is not identical to the cosmological
constant, in particular, it is not stable and eternal. Hence, it
is natural to seek nonstationary models of the current DE, too.

$f(R)$ gravity is one of those dynamical DE models in which the
Hilbert-Einstein action is modified and generalized by
incorporating a new phenomenological function of the Ricci scalar
$R$, $f(R)$. This theory provides a self-consistent and nontrivial
alternative to the $\lcdm$ model. It contains a new scalar degree
of freedom dubbed ``scalaron'' in Ref.~\cite{S80}.
The existence of this additional degree of freedom imposes a
number of conditions on viable functional forms of $f(R)$ for
$R\gg R_0$ and up to curvatures $R$ in the center of neutron
stars: \be |f(R)-R|\ll R,~~|f'(R)-1|\ll 1,~~Rf''(R)\ll 1~,
 \ee where the prime denotes the derivative with
respect to the argument $R$ and $R_0$ is a parameter of the order
of the present Ricci curvature. Furthermore, $f(R)$ should satisfy
the stability conditions:
\be f''(R)>0,~~f'(R)>0. \ee
Specific functional forms satisfying all these conditions have
been proposed in
Refs.~\cite{Hu:2007nk,Appleby:2007vb,Starobinsky:2007hu}.

In this paper, we have carried out numerical calculations of the
evolution of both a background space-time and density fluctuations
for the particular $f(R)$ model introduced in
Ref.~\cite{Starobinsky:2007hu}. As a result, we have found the
phantom boundary crossing at an intermediate redshift $z\lesssim
1$ for the background space-time metric and an anomalous behaviour
of the growth of density fluctuations.

\section{Background universe}%%%%%%%%%%%%%%%%%%%%%%%%%%%%%%%%%%%%%%%%%

%==================== Figure 1 a b ====================
\begin{figure}[t]
\centering
\includegraphics[width=80mm]{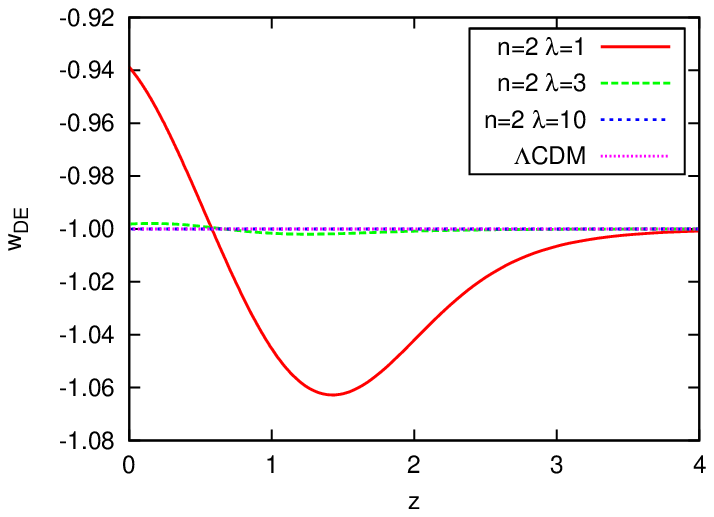}
\includegraphics[width=80mm]{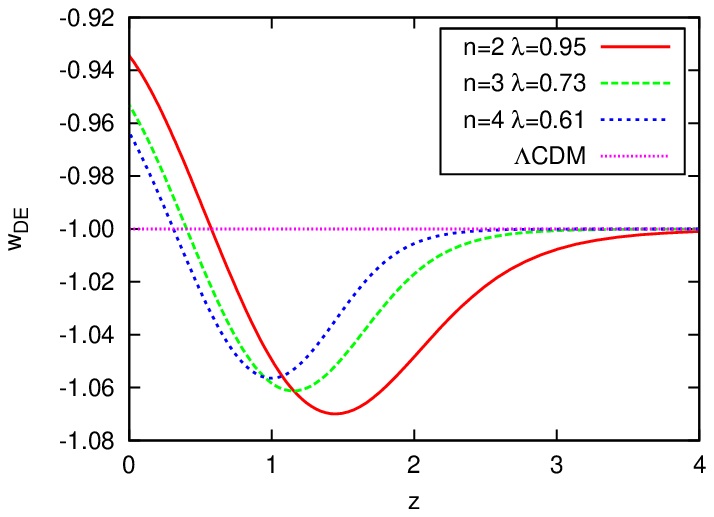}
\caption{
Evolution of the equation-of-state parameter $w_\DE(z)$ for the 
effective dark energy.
}
\label{fig:outbg_w}
\end{figure}
%==================== Figure 1 a b ====================

We adopt the following action with a three-parameter family of
$f(R)$ models:
\begin{align} 
&S= \f{1}{16\pi G}\int d^4x\g f(R)+ S_m, \label{S}\\
&f(R)=R+\lambda R_s\kk{\mk{1+\f{R^2}{R_s^2}}^{-n}-1}, \label{fR}
\end{align}
where $n,~\lambda$ and $R_s$
are model parameters and $S_m$ is
the action of the matter content.
We can derive field equations from the action \eqref{S} and
rewrite them as
\begin{align}
 R^{\mu}_{\nu}-\frac{1}{2}\delta^{\mu}_{\nu}R&=
-8\pi G\mk{T^{\mu}_{\nu (m)}+T^\mu_{\nu ({\rm DE})}}, \\
8\pi G T^\mu_{\nu (\DE)}&\equiv
(F-1)R^\mu_\nu-\frac{1}{2}(f-R)\delta^{\mu}_{\nu}
+(\nabla^\mu\nabla_\nu-\delta^{\mu}_{\nu}\square)F, \label{EMtensor}
\end{align}
where $F(R)\equiv f'(R)$.
Working in the
spatially flat Friedmann-Robertson-Walker (FRW) space-time with
the scale factor $a(t)$, we find
\begin{align}
3H^2&=8\pi G\rho-3(F-1)H^2+\frac{1}{2}(FR-f)-3H\dot F,\label{hubble}\\
2\dot{H}&=-8\pi G\rho -2(F-1)\dot{H}-\ddot{F}+H\dot{F},\label{hdot}
\end{align}
where $H\equiv \dot a/a$ is the Hubble parameter and $\rho$ is the energy density
of matter.

From Eq.~\eqref{EMtensor}, we can express the effective energy density
and pressure of dark energy as
\begin{align}
8\pi G\rho_{\rm DE}
&=-3H\dot{R}F' +3(H^2+\dot{H})(F-1)-\frac{1}{2}(f-R),\label{rhoDE}\\
8\pi G(\rho_{\rm DE}+P_{\rm DE})&=2\dot{H}(F-1)-H\dot{F}+\ddot{F}, \label{PDE}
\end{align}
respectively where $R=12H^2+6\dot{H}$. We define the DE equation
of state parameter $w_{\rm DE}$ by the ratio $w_{\rm DE}\equiv
P_{\rm DE}/\rho_{\rm DE}$, \be w_{\rm DE}\equiv\f{P_{\rm
DE}}{\rho_{\rm DE}}
=-1+\f{2\dot{H}(F-1)-H\dot{F}+\ddot{F}}{-3H\dot{R}F'
+3(H^2+\dot{H})(F-1)-(f-R)/2}. \ee

%==================== Table 1 ====================
\begin{table}[t]
\centering
\caption{$B(0)$ for various model parameters.}
\begin{tabular}{cl|c} \toprule
$n$ & $\lambda$ & $B(0)$ \\ \colrule
$2$ & $0.95$ & $2.09\times 10^{-1}$ \\
$2$ & $4$ & $9.36\times 10^{-4}$ \\
$2$ & $8$ & $6.07\times 10^{-5}$ \\
$3$ & $0.73$ & $1.86\times 10^{-1}$ \\
$3$ & $2$ & $1.34\times 10^{-3}$ \\
$3$ & $3$ & $1.35\times 10^{-4}$ \\
$4$ & $0.61$ & $1.73\times 10^{-1}$ \\
$4$ & $1$ & $1.33\times 10^{-2}$ \\ \bottomrule
\end{tabular} \label{B}
\end{table}
%==================== Table 1 ====================

 For the appropriate initial condition corresponding to the existence of
cosmic inflation in the past, $f-R$ acquires an asymptotically
constant value $f-R=-\lambda R_s$ at high redshift. In this regime
the evolution of the Universe is the same as that obtained from
the Einstein action with a cosmological constant
$\Lambda(\infty)=\lambda R_s/2$.

The late-time asymptotic de Sitter solution has a curvature
$R=R_1$ which is obtained as the maximal solution of the equation
$2f(R_1)=R_1f'(R_1)$, namely, \be \alpha(r)\equiv
r+2\lambda\kk{\f{1+(n+1)r^2}{(1+r^2)^{n+1}}-1}=0, \ee where
$r\equiv R_1/R_s$. It is obvious that the Minkowski space-time,
$r=0$, is one of the solutions. The stability condition of this
future de Sitter solution\cite{MSS88,Motohashi:2011prep},
$f'(R_1)>R_1f''(R_1)$, imposes the following constraint on $r$,
\be \beta(r) \equiv \f{(1+r^2)[(1+r^2)^{n+1}-2n\lambda
r]}{2n\lambda[(2n+1)r^2-1]}-r >0,\label{x1constraint} \ee which is
stronger than any other constraint discussed above. For each $n$,
we can find $r$, which marginally satisfies
Eq.~\eqref{x1constraint} and gives the minimal allowed value of
$\lambda$.  Numerically we find
$(n,r_{\min},\lambda_{\min})=(2,1.267,0.9440),(3,1.041,0.7259)$,
and $(4,0.9032,0.6081)$ for $n\le 4$.

We numerically solve the evolution equation \eqref{hdot} using
Eq.~\eqref{hubble} to check the numerical accuracy and taking
$t_i$ as the moment of time when the matter density parameter was
$\Omega_i=16\pi G\rho_i/(16\pi G\rho_i+\lambda R_s)=0.998$. We
determine the current epoch by the requirement that the value of
$\Omega$ takes the observed central value $\Omega_0=0.27$. Then
$R_s$ is fixed in such a way that the current Hubble parameter
$H_0=72~$km/s/Mpc is reproduced.

Figure \ref{fig:outbg_w} depicts the evolution of $w_{\rm DE}$ as
a function of redshift $z$ \cite{Motohashi:2010tb}. Phantom
crossing is manifest there. As expected, $w_{\rm DE}$ approaches
$-1=\text{constant}$ as we increase $\lambda$ for fixed $n$. For
$\lambda=\lambda_{\min}$, deviations from $w_{\rm DE}=-1$ are
observed at $\sim  5\%$ level in both directions for $z \lesssim
2$ independently of $n$. Such behaviour of $w_{\rm DE}$ is well
admitted by all the most recent observational data\cite{WMAP7}.
The average value of $w_{\rm DE}$ over the interval $0 \le z \le
1$ to which all BAO and most of the SN data refer is very close to
$-1$. Moreover, in this range
the behaviour of $w_{\rm DE}$ for $\lambda=\lambda_{\min}$
is well fitted by the CPL fit\cite{CPL} $w_{\rm
DE}(z)=w_0+w_az/(1+z)$ with $(n,w_0,w_a)=$ $(2,-0.92,-0.23)$,
$(3,-0.94,-0.22)$, and $(4,-0.96,-0.21)$, respectively. These
values of $w_0$ and $w_a$ lie very close to the center of the
$68\%$ and $95\%$ CL ellipses for all combined data in Fig. 13 of
Ref.~\cite{WMAP7}. This phantom crossing is not peculiar to the
specific choice of the function \eqref{fR} but a generic one for
models which satisfy the stability condition $F'>0$.

We have also calculated the quantity
$B(z) = (f''/f')(dR/d\ln H)$
at the present time.
The results are presented in Table \ref{B}.

\section{Density fluctuations}%%%%%%%%%%%%%%%%%%%%%%%%%%%%%%%%%%%%%%%%%

Armed with the evolution of the Universe background, we proceed to
the investigation of the evolution of matter density fluctuations,
$\d$, in $f(R)$ gravity. In the subhorizon limit, the evolution equation
is derived as
\cite{Zhang:2005vt,Tsujikawa:2007gd}
\begin{align} 
&\ddot \d + 2H\dot \d - 4\pi G_\eff \rho \d = 0,\label{de1} \\
&G_\eff=\frac{G}{F}\frac{1+4\frac{k^2}{a^2}\frac{F'}{F}}{1+3\frac{k^2}{a^2}\frac{F'}{F}}. 
\end{align}
Equation~\eqref{de1} reduces to the correct evolution equation for
all wavenumbers for the $\Lambda$CDM model in the Einstein gravity
where $F=1$. The time and $k$-dependence of the effective
gravitational constant $G_\eff$ changes the evolution of
density fluctuations.

\begin{figure}[t]
\centering
\includegraphics[width=80mm]{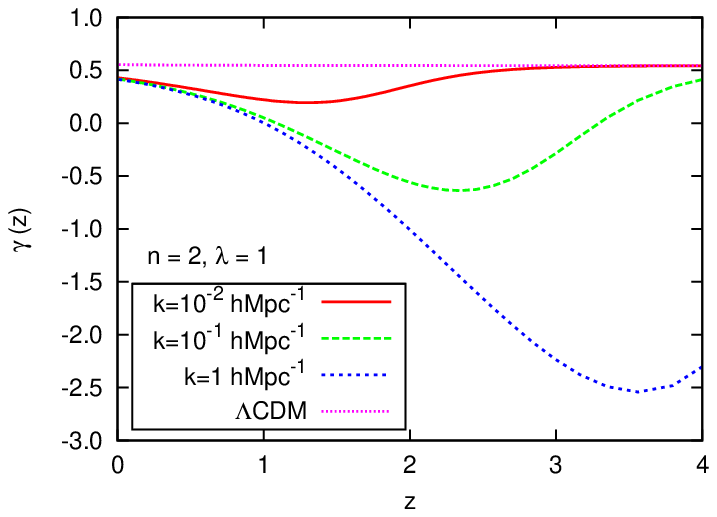}
\includegraphics[width=80mm]{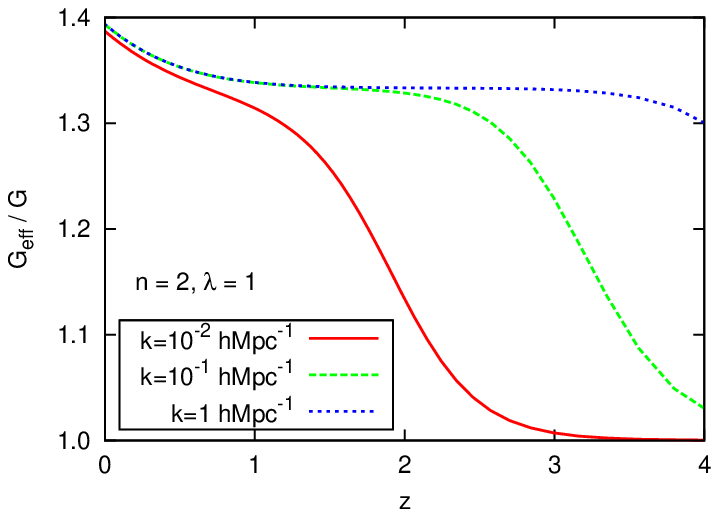}
\caption{
Evolutions of the growth index $\gamma(z)$ and the effective gravitational constant $G_{\eff}(z)$ for $n=2$ and $\lambda=1$.
}
\label{fig:gG}
\end{figure}

First, let us consider the gravitational 
growth index,
$\gamma(z)$, which is an important quantity helping to distinguish
different modified theories of gravity. It is defined through 
\be
 \f{d\ln\delta}{d\ln a}=\Omega_m(z)^{\gamma(z)},~~~\text{or}~~~
 \gamma(z)=\f{\log\mk{\f{\dot{\delta}}{H\delta}}}{\log\Omega_m}.
\ee 
In the standard $\lcdm$ model, it takes a practically constant
value $\gamma\cong 0.55$. However, it evolves with time in
modified gravity theories in general. We also note that
$\gamma(z)$ has a nontrivial $k$-dependence in $f(R)$ gravity,
since density fluctuations with different wavenumbers evolve
differently. Therefore, this quantity is a useful diagnostic to
distinguish DE in modified gravity from the $\lcdm$ model in the
Einstein gravity. We present the evolution of $\gamma(z)$ together
with that of $G_{\eff}/G$ for different values of $k$ in
Fig.~\ref{fig:gG}. $\gamma(z)$ takes a constant value identical to
the $\lcdm$ model in the early high-redshift regime because $f(R)$
gravity is indistinguishable from the Einstein gravity plus a
positive cosmological constant then. It gradually decreases with
time, reaches a minimum, and then increases again towards the
present epoch. Current constraints on the growth
index\cite{Rapetti:2009ri} are not strong enough to detect any
deviation from the $\lcdm$ model and/or to obtain new bounds on
$f(R)$ DE models, but future observations may reveal its time and
wavenumber dependences.

%==================== Figure 2 a b ====================
%==================== Figure 2 a b ====================

\begin{figure}[t]
\centering
\includegraphics[width=80mm]{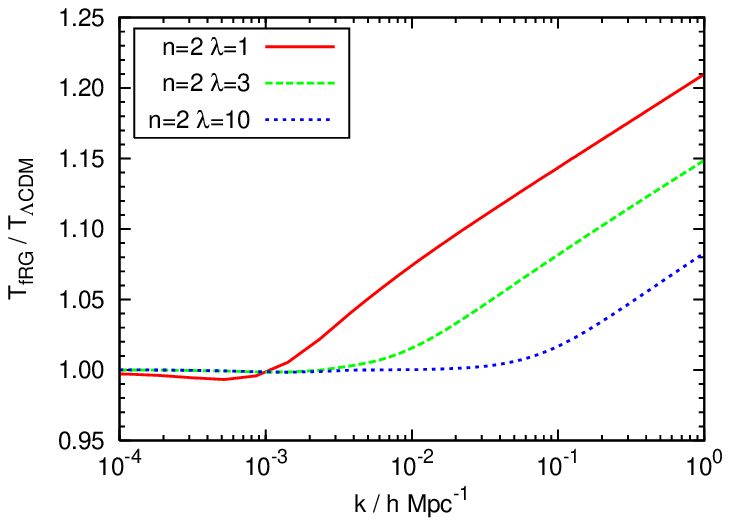}
\includegraphics[width=80mm]{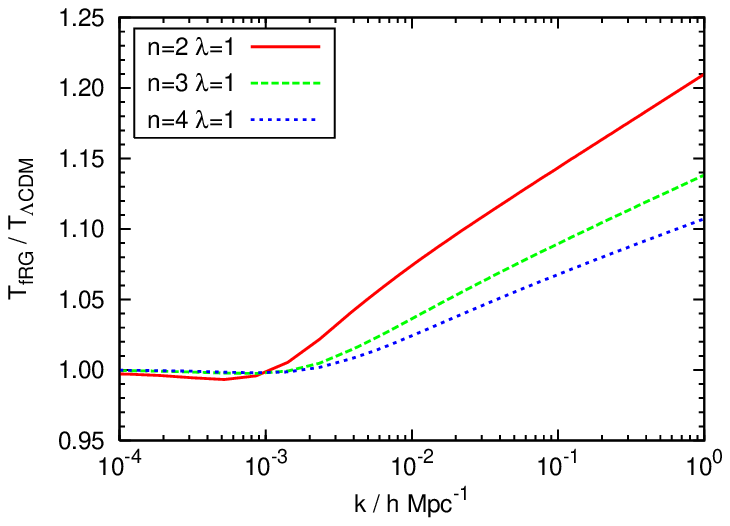}
\caption{
Additional transfer function $C(k)$ in $f(R)$ gravity.
}
\label{fig:TF}
\end{figure}

Second, we focus on the additional late-time transfer function for
linear matter perturbations arising in $f(R)$ gravity. 
The evolution equation 
\eqref{de1} can be solved analytically in the
high-curvature regime when the scale factor evolves as
$a(t)\propto t^{2/3}$ and $F$ takes the asymptotic form 
$F\simeq 1-2n\lambda \mk{R/R_s}^{-2n-1}\equiv 1-\mk{R/R_c}^{-N-1}$,
where
$N=2n$ and $R_c=R_s(2n\lambda)^{1/(2n+1)}$.
The two independent solutions of 
Eq.~\eqref{de1} in this regime read
\begin{align}
&\delta_{\veck}(t)=\delta_{i\veck}\left(\frac{t}{t_i}\right)^{\frac{-1\pm
5}{6}} \nonumber \\
&\times \,_2F_1\left(\frac{\pm 5-\sqrt{33}}{4(3N+4)}, \frac{\pm
5+\sqrt{33}}{4(3N+4)};1\pm\frac{5}{2(3N+4)};
-3\frac{(N+1)k^2}{a_i^2R_c^2}\left(\frac{t}{t_i}\right)^{2N+8/3}
\right)  \label{as}
\end{align}
in terms of the hypergeometric function\cite{Motohashi:2009qn}.
Hereafter, we consider the upper
sign solution only, because the other solution corresponds to the
decaying mode and is singular at $t\to 0$.
Then, the solution behaves as
\be
  \delta_{\veck}(t)\xrightarrow[]{t\to 0} \delta_{i\veck}
\left(\frac{t}{t_i}\right)^{\frac{2}{3}}
~~{\rm and}~~
\delta_{\veck}(t)\xrightarrow[]{t\to \infty}
\delta_{i\veck}C(k)\left(\frac{t}{t_i}\right)^{\frac{-1+\sqrt{33}}{6}},
 \label{limiting}
\ee
respectively.  The transfer function, $C(k)$, is given by
\begin{align}
C(k)&=\f{\Gamma\left(1+\frac{5}{2(3N+4)}\right)
\Gamma\left(\frac{\sqrt{33}}{2(3N+4)}\right)}
{\Gamma\left(1+\frac{5+\sqrt{33}}{4(3N+4)}\right)
\Gamma\left(\frac{5+\sqrt{33}}{4(3N+4)}\right)}
\left[\frac{3(N+1)k^2}{a_i^2R_c}
\left(\frac{3R_ct_i^2}{4}\right)^{N+2}\right]
^{\frac{-5+\sqrt{33}}{4(3N+4)}},
\label{transfer}
\end{align}
where 
$t_i=2/3\sqrt{6/\lambda R_s}\sinh^{-1}\sqrt{(1-\Omega_i)/\Omega_i}$. 
We have confirmed
the additional transfer function $C(k)$
numerically as Fig.~\ref{fig:TF}.

%==================== Figure 3 a b ====================
%==================== Figure 3 a b ====================

%==================== Figure 4 a b ====================
\begin{figure}[t]
\centering 
\includegraphics[width=80mm]{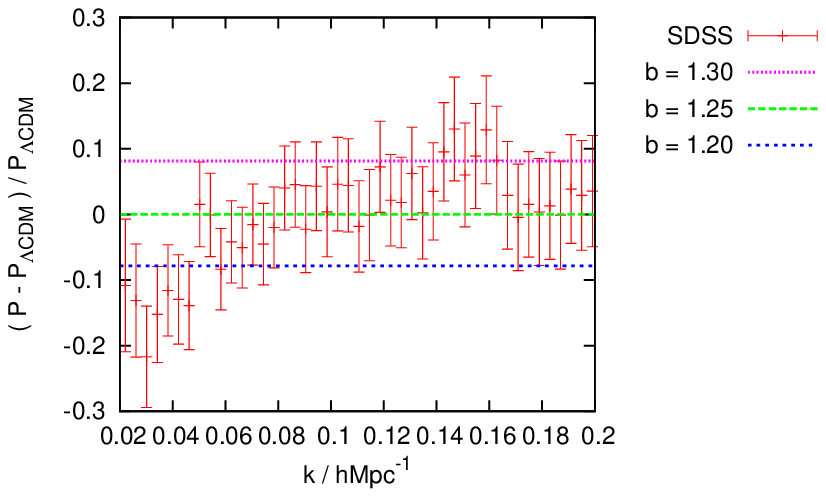}
\includegraphics[width=80mm]{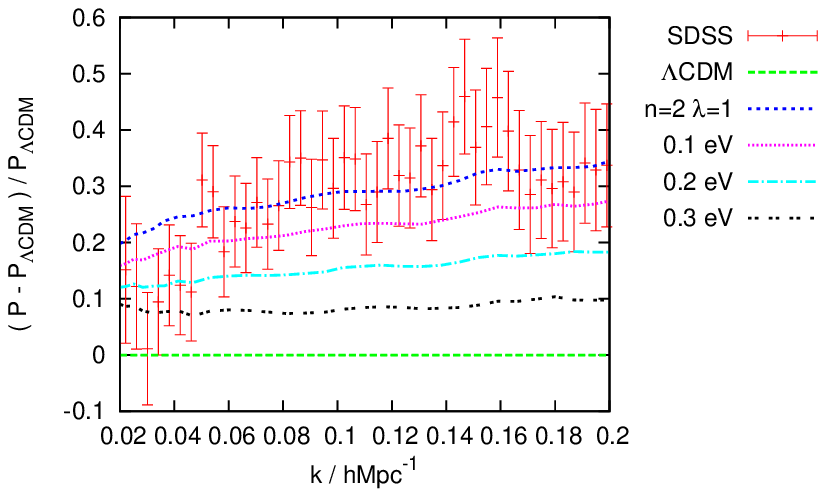}
\caption{ Power spectrum with SDSS data. 
Left: Power spectrum in the $\lcdm$ model normalized by $P_\lcdm(b=1.25)$.
Right: Power spectrum in $f(R)$ gravity for $n=2$ and $\lambda=1$
with total neutrino mass up to $0.3$eV normalized by $P_\lcdm(b=1.1)$.
} 
\label{fig:psr}
\end{figure}
%==================== Figure 4 a b ====================

Finally, we proceed to compare the results of our numerical
calculations for matter power spectrum with observational data
from SDSS DR7\cite{Reid:2009xm}. The left panel of
Fig.~\ref{fig:psr} shows the fitting of the matter power spectrum
for the $\lcdm$ model. The vertical axis is the ratio of the power
spectrum normalized by the power spectrum in $\lcdm$ model with
bias parameter $b=1.25$. The data with error bars are LRG samples
in SDSS DR7. As far as we use a constant bias, we see that
observational data are increasing with wavenumber compared with
the theoretical expectation line in the $\lcdm$ model.
Furthermore, if we include massive neutrinos, the fitting becomes
worse because their free streaming suppress the matter power
spectrum. We have found that $f(R)$ gravity enhances the matter
power spectrum and produces a better fit to the SDSS data than the
$\lcdm$ model without introducing a scale dependent bias or a
nonlinear bias, see the right panel of Fig.~\ref{fig:psr}. There
is also a possibility to admit a larger total sum of neutrino
restmasses compared to the $\lcdm$ model since the anomalous
growth of perturbations partially counteracts their suppression by
free streaming of massive neutrinos~\cite{Motohashi:2010sj}.

\section{Conclusion}%%%%%%%%%%%%%%%%%%%%%%%%%%%%%%%%%%%%%%%%%

In this paper, we have numerically calculated the evolution of
both a homogeneous isotropic background and matter density
fluctuations in a viable $f(R)$ DE model based on the specific
functional form proposed in Ref.~\cite{Starobinsky:2007hu}. We
have found that viable $f(R)$ gravity models of the present DE and
accelerated expansion of the Universe generically exhibit phantom
behaviour during the matter-dominated stage with crossing of the
phantom boundary $w_{\rm DE}=-1$ at redshifts $z\lesssim 1$. More
exactly, this behaviour is characteristic for all $f(R)$ DE models
that have $f''(R)>0$ and a stable future de Sitter epoch and which
approach the Einstein gravity sufficiently fast for $R\gg R_0$,
under the condition that the gravitational constant $G$ in the
Einsteinian representation of the field equations (4) is
normalized to its value measured in laboratory experiments (i.e.,
for $R\gg R_0$, too). The predicted time evolution of $w_{\rm DE}$
has qualitatively the same behaviour as that has recently obtained
from observational data\cite{Shafieloo:2009ti}. However, it is
important that the condition of stability, or even metastability,
of the future de Sitter epoch strongly restricts the possible
deviation of $w_{\rm DE}$ from $-1$ by several percents in these
models. Thus, the DE phantomness should be small, if it exists at
all, that agrees well with the present observational data.

As for the density fluctuations, we have also investigated the
growth index $\gamma(k,z)$ of density fluctuations and have
presented an explanation of its anomalous evolution in terms of
the time dependence of $G_{\eff}$. Note that this evolution is
characteristic for all $f(R)$ models in which the scalar particle
(scalaron) becomes relativistic ($k^2>m_s^2(R)a^2$) at recent
redshifts. Since $\gamma$ has a characteristic time and wavenumber
dependence, future detailed observations may yield useful
information on the validity of $f(R)$ gravity through this
quantity, although current constraints have been obtained assuming
that it is constant both in time and in
wavenumber\cite{Rapetti:2009ri}.

We have also numerically confirmed a shift in the power spectrum
index for larger wavenumbers which exceed the scalaron mass during
the matter-dominated epoch\cite{Motohashi:2009qn}, while for
smaller wavenumbers, fluctuations have the same amplitude as in
the $\lcdm$ model. Once more, the future de Sitter epoch stability
condition bounds a possible increase in density fluctuations for
cluster scales (compared with the $\lcdm$ model) by $\sim 40\%$
for $n\ge 2$. This enhancement in matter power spectrum can
explain the observational data from SDSS and allows for some increase
in the total sum of neutrino restmasses, as compared to the
standard $\lcdm$ model.

\section*{Acknowledgments}%%%%%%%%%%%%%%%%%%%%%%%%%%%%%%%%%%%%%%%%%

AS acknowledges RESCEU hospitality as a visiting professor. He was
also partially supported by the grant RFBR 11-02-00643 and by the
Scientific Programme ``Astronomy'' of the Russian Academy of
Sciences. This work was supported in part by JSPS Research
Fellowships for Young Scientists (HM), JSPS Grant-in-Aid for
Scientific Research No.\ 19340054 (JY), Grant-in-Aid for
Scientific Research on Innovative Areas No. 21111006 (JY), JSPS
Core-to-Core program ``International Research Network on Dark
Energy'', and Global COE Program ``the Physical Sciences
Frontier'', MEXT, Japan.

\end{document}